\documentclass[prl,twocolumn,showpacs,amsmath,amssymb,superscriptaddress]{revtex4}

\usepackage[dvips]{graphicx}% Include figure files

\usepackage{graphics}% Include figure files

\usepackage{dcolumn}% Align table columns on decimal point

\usepackage{bm}% bold math

\usepackage[usenames]{color}

\begin{document}

\title{ Voltage control of magnetocrystalline anisotropy \\ in ferromagnetic -- semiconductor/piezoelectric hybrid structures }

\author{A.~W.~Rushforth}
\altaffiliation{Corresponding author}
\email{Andrew.Rushforth@Nottingham.ac.uk}

\affiliation{School of Physics and Astronomy, University of Nottingham, Nottingham NG7 2RD, United Kingdom}

\author{E. De Ranieri}
\affiliation{Hitachi Cambridge Laboratory, Cambridge CB3 0HE, United Kingdom}

\author{J.~Zemen}
\affiliation{Institute of Physics ASCR v.v.i., Cukrovarnick\'a 10, 162 53 Praha 6, Czech Republic}

\author{J.~Wunderlich}
\affiliation{Hitachi Cambridge Laboratory, Cambridge CB3 0HE, United Kingdom}
\affiliation{Institute of Physics ASCR v.v.i., Cukrovarnick\'a 10, 162 53 Praha 6, Czech Republic}

\author{K.~W.~Edmonds}
\affiliation{School of Physics and Astronomy, University of Nottingham, Nottingham NG7 2RD, United Kingdom}

\author{C.~S.~King}
\affiliation{School of Physics and Astronomy, University of Nottingham, Nottingham NG7 2RD, United Kingdom}

\author{E.~Ahmad}
\affiliation{School of Physics and Astronomy, University of Nottingham, Nottingham NG7 2RD, United Kingdom}

\author{R.~P.~Campion}
\affiliation{School of Physics and Astronomy, University of Nottingham, Nottingham NG7 2RD, United Kingdom}

\author{C.~T.~Foxon}
\affiliation{School of Physics and Astronomy, University of Nottingham, Nottingham NG7 2RD, United Kingdom}

\author{B.~L.~Gallagher}
\affiliation{School of Physics and Astronomy, University of Nottingham, Nottingham NG7 2RD, United Kingdom}

\author{K. V\'yborn\'y}
\affiliation{Institute of Physics ASCR v.v.i., Cukrovarnick\'a 10, 162 53 Praha 6, Czech Republic}

\author{J. Ku\v{c}era}
\affiliation{Institute of Physics ASCR v.v.i., Cukrovarnick\'a 10, 162 53 Praha 6, Czech Republic}

\author{T.~Jungwirth}
\affiliation{Institute of Physics ASCR v.v.i., Cukrovarnick\'a 10, 162 53 Praha 6, Czech Republic}
\affiliation{School of Physics and Astronomy, University of Nottingham, Nottingham NG7 2RD, United Kingdom}

\date{\today}

\begin{abstract}

We demonstrate dynamic voltage control of the magnetic anisotropy of a (Ga,Mn)As device bonded to a piezoelectric transducer. The application of a uniaxial strain leads to a large reorientation of the magnetic easy axis which is detected by measuring longitudinal and transverse anisotropic magnetoresistance coefficients. Calculations based on the mean-field kinetic-exchange model of (Ga,Mn)As provide microscopic understanding of the measured effect. Electrically induced magnetization switching and detection of unconventional crystalline components of the anisotropic magnetoresistance are presented, illustrating the generic utility of the piezo voltage control to provide new device functionalities and in the research of micromagnetic and magnetotransport phenomena in diluted magnetic semiconductors.

\end{abstract}


\maketitle

%\protect\newpage

The control of magnetic properties of ferromagnetic materials by electrical means is an important prerequisite for successful implementation of spintronics in information processing technologies, and for major advancements in sensor and transducer applications. The material class of (III,Mn)V dilute magnetic semiconductors (DMSs) plays a major role in this research because it combines ferromagnetic and semiconducting properties in one physical system and because of the potential compatibility with modern microelectronic technologies. The intense research into DMS systems has led to several successful demonstrations of direct gate  control of ferromagnetism via carrier redistribution \cite{Chiba:2003_a,Chiba:2006_b}. However, the high carrier-doping levels ($>\sim1$\%) necessary to achieve ferromagnetism render the direct gating efficiency in these semiconductor materials relatively low.

While unfavorable for the direct gating, the large concentrations (Fermi energies) of the spin polarized holes that mediate ferromagnetic coupling between the Mn local moments produce large magnetic stiffness, resulting in a mean-field like magnetization and macroscopic single-domain characteristics of these dilute moment ferromagnets. At the same time, magnetocrystalline anisotropies derived from spin-orbit coupling effects in the hole valence bands remain large, even at these high hole concentrations. This leads to the sensitivity of magnetic easy-axes orientations to strains as small as $10^{-4}$ \cite{Wunderlich:2007_c,Wenisch:2007_a}. So far the strain effects have been controlled by lattice parameter engineering  during growth \cite{Dietl:2001_b,Abolfath:2001_a} or through post growth lithography \cite{Wenisch:2007_a,Humpfner:2006_a,Wunderlich:2007_c}.

%There is also a remarkably favorable aspect in the physics of DMSs 
%stemming from large densities, combined with large spin-polarization 
%and spin-orbit coupling, of the holes that mediate ferromagnetic 
%coupling between dilute Mn moments. The holes provide large magnetic 
%stiffness and anisotropy energy resulting in macroscopic single-domain %behavior of these ferromagnets. The magnetic anisotropies are 
%sensitive to strains as small as $10^{-4}$ \cite{Wunderlich:2007
%_c,Wenisch:2007_a}. So far the strain effects have been controlled by %lattice parameter engineering  during growth \cite{Dietl:2001
%_b,Abolfath:2001_a} or through post growth lithography \cite
%{Wenisch:2007_a,Humpfner:2006_a,Wunderlich:2007_c}. 

It has been demonstrated that sizeable strains can be induced in GaAs structures using a piezoelectric transducer \cite{Shayegan:2003_a}. Here we utilize this technique to demonstrate the dynamic voltage control via strain of the magnetic anisotropy in a (Ga,Mn)As device bonded to a piezo-transducer.
We demonstrate that the favorable micromagnetic characteristics of DMSs and the relatively simple band structure
allow for a microscopic description of these effects on an unprecedented level of accuracy compared to metal ferromagnet/piezoelectric devices \cite{Kim:2003_a,Lee:2003_b,Botters:2006_a,Boukari:2007_a}.
Finally we discuss the realization of electrically induced magnetization switching and of the detection of unconventional crystalline components of the anisotropic magnetoresistance (AMR). These are two examples illustrating the generic utility of the piezo voltage control to provide new device functionalities and in the research of micromagnetic and magnetotransport phenomena in DMSs.

The 25~nm thick Ga$_{0.94}$Mn$_{0.06}$As epilayer was grown by low-temperature molecular-beam-epitaxy on GaAs substrate and buffer layers (see \cite{Campion:2003_a} for details). The material is under compressive in-plane strain of $\sim$ 3 x 10$^{-3}$ \cite{Zhao:2005_a} due to the lattice mismatch with the GaAs. From SQUID magnetometry on an unstrained sample the magnetic easy axis is in-plane in a direction determined by competition between biaxial [100]/[010] and uniaxial [1$\bar{1}$0] anisotropies. The uniaxial term dominates above 10~K. At 50~K the cubic and uniaxial anisotropy constants determined from hard axis magnetisation curves are $K_c=85$~Jm$^{-3}$ and $K_u=261$~Jm$^{-3}$ $\pm$ ~20\%.

A (Ga,Mn)As Hall bar, fabricated by optical lithography, and orientated along the [1$\bar{1}$0] direction, was bonded to the PZT piezo-transducer using a two-component epoxy after thinning the substrate to $150 \pm 10$~$\mu$m by chemical etching. The stressor was slightly misaligned so that a positive/negative voltage produces a uniaxial tensile/compressive strain at $\approx -10^{\circ}$ to the [1$\bar{1}$0] direction.

The induced strain was measured by strain gauges, aligned along the [1$\bar{1}$0] and [110] directions, mounted on a second piece of $150 \pm 10$~$\mu$m thick wafer bonded to the piezo-stressor. Differential thermal contraction of GaAs and PZT on cooling to 50~K produces a measured biaxial, in plane, tensile strain at zero bias of ~$10^{-3}$ and a uniaxial strain estimated to be of the order of $\sim 10^{-4}$ \cite{Habib:2007_a} which could not be accurately measured. At 50~K, the magnitude of the additional strain for a piezo voltage of $\pm 150$~V is approximately $2\times 10^{-4}$.

In this study the orientation of the in-plane magnetisation of the (Ga,Mn)As Hall bar was determined from the measured longitudinal and transverse AMR. To a good approximation ($\approx 10\%$), these are given by $\Delta\rho_{xx}/\rho_{av}= C\cos2\phi$ and $\rho_{xy}/\rho_{av}=C\sin2\phi$, where $\phi$ is the angle between the magnetisation direction and the Hall bar (current) direction \cite{Rushforth:2007_a}.

Figure~1 shows magnetoresistance measurements at 50~K for external magnetic field sweeps at constant field angle $\theta$ measured from the Hall bar direction. The strongly $\theta$-dependent low-field magnetoresistance, which saturates at higher field is due to AMR, i.e., to magnetisation rotations. We have subtracted an isotropic, $\theta$-independent magnetoresistance contribution, approximated as linear, from the measured longitudinal resistances.

\begin{figure}[t]
%\vspace{1cm}
\hspace{1cm}\includegraphics[width=1.0\columnwidth,angle=0]{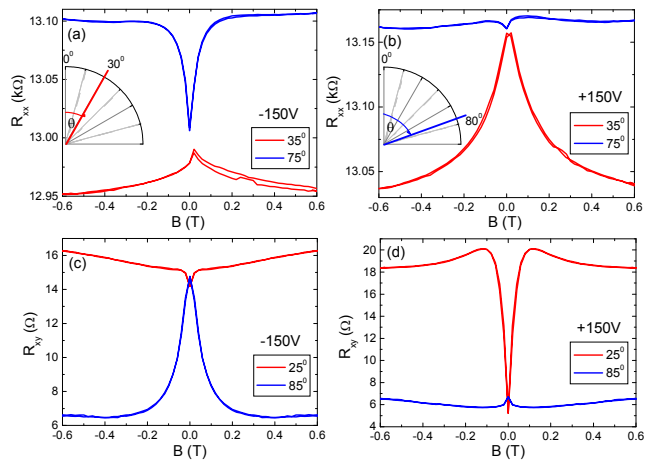}

%\vspace*{-2cm}
%\caption{\textbf{Figure1}}
\caption{The longitudinal resistances, R$_{xx}$ ((a) and (b)) and the
transverse resistances R$_{xy}$ ((c) and (d)) as a function of
magnetic field for angles close to the easy axes (30$^{\circ}$ at -150
~V and 80$^{\circ}$ at +150~V). The curves close to the easy
axes in each case are relatively flat as a function of field,
indicating small rotation of the angle of the magnetisation. T=50~K}
\label{f1}
\end{figure}

When the external field is close to the magnetic easy axis, the measured resistances at saturation and remanence should be almost the same and a significant magnetoresistance due to rotation of the magnetisation can only be present at very low applied fields. For external fields away from the easy axis, large magnetoresistances corresponding to large rotations of the magnetisation orientation are present.  This enables us to determine the easy axis directions within $\pm 5^{\circ}$.

The effect of the piezo-stressor is clearly apparent in Figure~1. At 50K, SQUID measurements show that the magnetic easy axis is oriented along the [1$\bar{1}$0]- direction for the as-grown (Ga,Mn)As wafer, consistent with $|K_c| < |K_u|$. The easy axis for the Hall bar bonded to the stressor rotates to an angle $\phi=65^{\circ}$ upon cooling to 50~K due to a uniaxial strain induced by anisotropic thermal contraction of the piezo stressor \cite{Habib:2007_a}. Application of a bias of +150~V to the stressor causes the easy axis to rotate further to $\phi=80^{\circ}$ while for -150~V it rotates in the opposite sense to $\phi=30^{\circ}$. This directly demonstrates electric field control of the magnetic anisotropy in our (Ga,Mn)As/PZT hybrid system.

The magnetic anisotropy for our system can be described phenomenologically by an energy functional  $E(\hat{M})=-K_c/4 \sin^2 2\phi+K_u \sin^2\phi+K_u^{\prime}\sin^2(\phi+\phi_{0})$, where the last term, with $\phi_{0}\approx10^{\circ}$, is due to the misaligned stressor. The observed behaviour is then consistent with the (Ga,Mn)As being in tensile strain along the axis of the stressor on cool down and applied positive (negative) voltage increasing (decreasing) this strain. Note that the misalignment allows smooth rotation of a single easy axis in the experimentally accessible voltage range.

We now calculate the expected magnetic anisotropy characteristics of the studied (Ga,Mn)As/PZT system by combining the six-band ${\bf k}\cdot{\bf p}$ description of the GaAs host valence band with the kinetic-exchange model of the coupling to the local Mn$_{\rm Ga}$ $d^5$-moments \cite{Dietl:2001_b,Abolfath:2001_a}. This approach is well-suited to the description of spin-orbit coupling phenomena near the top of the valence band whose spectral composition and related symmetries are dominated by the $p$-orbitals of the As sublattice, and also provides straightforward means of incorporating lattice strains \cite{Dietl:2001_b,Abolfath:2001_a,Wunderlich:2007_c}.

\begin{figure}[h]

%\vspace{1cm}
\hspace*{0cm}\includegraphics[width=0.8\columnwidth,angle=0]{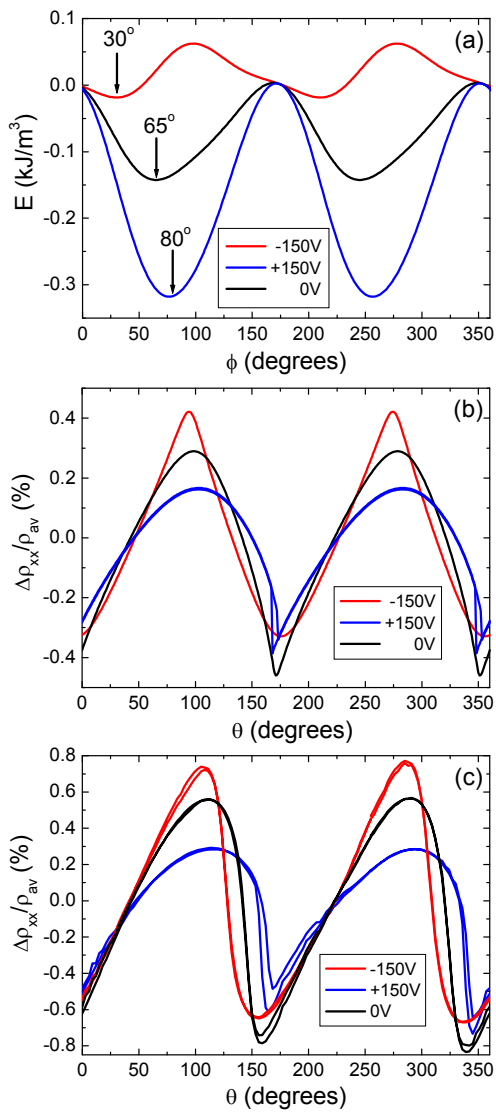}
%\vspace{-0.1cm}
%\caption{\textbf{Figure 2}}
\caption{(a) The microscopic $E(\hat{M})$ curves for the three
piezo voltages. $\phi$ is the angle of the magnetisation with respect
to the Hall bar. (b) The longitudinal AMR from theory calculations
with a non saturating magnetic field of 20~mT rotated in the plane of
the film (c) The experimental AMR curves with a field of 40~mT rotated
in the plane of the film. $\rho_{av}$ is the $\rho_{xx}$ averaged over
360$^{\circ}$ in the low field regime. $\theta$ is the angle of the
magnetic field with respect to the Hall bar. T=50~K}
\label{f2}
\end{figure}

Due to the presence of unintentional compensating defects in (Ga,Mn)As films, the concentrations of ferromagnetically ordered Mn local moments and holes cannot be accurately controlled during growth or determined post growth \cite{Jungwirth:2005_a}. We therefore consider in our analysis uncompensated Mn$_{\rm Ga}$ moment concentrations within an interval $x=3-5\%$ which safely contain the expected value of x. The magnetocrystalline anisotropy constants depend strongly on the local moment density and the hole compensation ratio $p/N_{\rm Mn}$, where p is the hole density and $N_{\rm Mn}$ is the concentration of Mn ions. For fixed $p$ and $N_{\rm Mn}$, the cubic term $K_c$,  calculated without adjustable parameters, agrees with the measured 50~K value for $p/N_{\rm Mn}=0.6-0.4$ for $x=3-5\%$ in good agreement with the estimated compensation ratio in our as-grown material \cite{Jungwirth:2005_a}.

The origin of the uniaxial anisotropy term in bare (Ga,Mn)As wafers is not known, but it can be modelled \cite{Sawicki:2004_a,Wunderlich:2007_c} by introducing a shear strain $e_{int}$ along the [1$\bar{1}$0] axis. For $p/N_{\rm Mn}=0.6-0.4$ we obtain the experimental $T=50$~K value of $K_u$ for compressive shear strain $e_{int}=3-2\times 10^{-4}$ within the considered range of $x$'s.

The calculations reproduce the measured 0~V easy axis for a tensile strain of $e_{str}=6-4\times 10^{-4}$, along the stressor axis and the experimental easy axes for $\pm 150$~V are obtained by increasing/decreasing the $e_{str}$ strain by $3-2\times 10^{-4}$. These \textit{changes} in strain agree with the measured values for $\pm 150$~V, and the 0~V strain due to differential contraction is of the expected order. The resulting microscopic $E(\hat{M})$ curves for the three voltages are shown in Figure~2(a).

The magnetoresistance calculated microscopically from the same band structure model combined with Boltzmann transport theory \cite{Rushforth:2007_a} gives AMR at saturation of the same sign and comparable magnitude to the experiment if we assume the above compensation ratios. This allows us to microscopically simulate AMR measurements assuming the single domain behaviour. In Figure~2(b) we show the results of simulations and in Figure~2(c) experimental data for the situation where a magnetic field of magnitude smaller than the saturation field is rotated in the plane of the (Ga,Mn)As epilayer. Both theory and experiment show that these AMR traces are no longer sinusoidal since the magnetisation does not track the applied rotating field. Ranges of magnetic field angles $\theta$ for which resistance is more slowly varying correspond to angles close to the easy axis, where the magnetisation vector lags behind $\theta$.  On the other hand rotation around the hard axis is more abrupt, and in this region the AMR can develop hysteretic features whose widths increase with decreasing magnitude of the rotating field.  At +150~V the hard axis is close to the Hall bar axis resulting in sharper minima than maxima in the corresponding experimental and theoretical AMR traces, while the trend is clearly opposite for the -150~V bias data, consistent with the easy axis directions obtained from the field sweep measurements.

\begin{figure}[h]

%\vspace{1cm}
\hspace*{0cm}\includegraphics[width=0.8\columnwidth,angle=0]{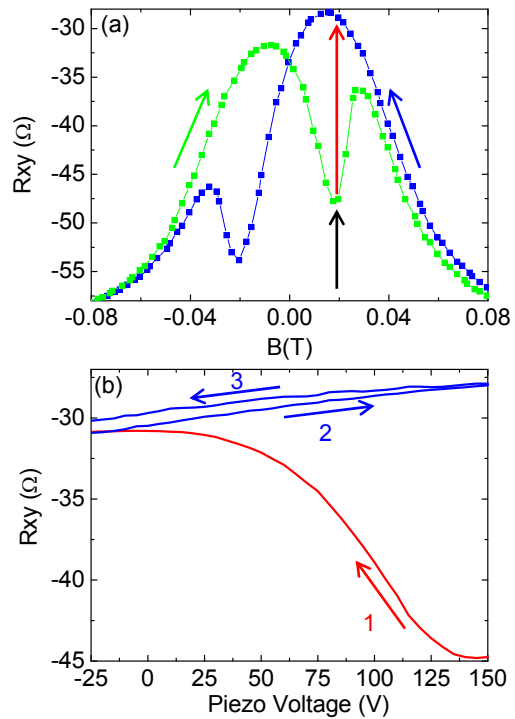}
%\vspace{-0.1cm}
%\caption{\textbf{Figure 3}}
\caption{(a) Low field magnetic hysteresis curve
at +150~V. The field is swept from saturating negative field at 165$^
{\circ}$ to the position show by the black arrow. Then (b) the piezo
voltage is swept inducing a rotation of the angle of the
magnetisation, indicated by the red arrows. Numbered arrows represent
the order and direction of the voltage sweeps. T=30~K.}
\label{f3}
\end{figure}

Having established a microscopic understanding of the control of the magnetic anisotropy in the (Ga,Mn)As/PZT hybrid system we now proceed with the demonstration of an electrically induced magnetisation switching. The bias-dependent hysteresis loops which allow for such a reversal process are shown in Figure~3(a). With the piezo voltage at +150V, the initial magnetisation state is prepared by sweeping the external magnetic field from negative saturating field at 165$^{\circ}$ to the position shown by the black arrow. This causes the magnetisation to rotate from 165$^{\circ}$ to 260$^{\circ}$, at B=0~T (i.e. along the easy axis at +150~V), then to 275$^{\circ}$ for the small positive field of approximately 18~mT (marked by the black arrow). Then, with the external magnetic field held constant the piezo voltage is swept (Figure 3(b)) and the magnetisation rotates from 275$^{\circ}$ to 25$^{\circ}$ (i.e. close to the easy axis for -150~V) resulting in a change of $R_{xy}$ as shown by the red arrows. This sequence switches the magnetisation from the 4th to the 1st quadrant, where it remains for subsequent voltage sweeps. The magnetisation can be switched back again by reversing the sequence, with the magnetic field set to the opposite polarity.

Finally we report on the detection of an unconventional crystalline component of the AMR allowed by the piezo voltage control. The AMR in (Ga,Mn)As is known to consist of a non-crystalline component, reflecting the symmetry breaking imposed by a preferred current direction, and crystalline terms reflecting the underlying crystal symmetry. The crystalline terms typically represent 10\% of the total AMR in 25nm (Ga,Mn)As layers \cite{Rushforth:2007_a}. Figure 4 shows the change in the longitudinal $\Delta\rho_{xx}/\rho_{av}$ and transverse $\Delta\rho_{xy}/\rho_{av}$ components of the AMR for piezo voltages of $\pm 150$~V. $\Delta\rho_{xx}=\rho_{xx}-\rho_{av}$, and $\rho_{av}$ is the average of $\rho_{xx}$ over 360$^{o}$ in the plane. The distortion of the lattice by the piezo transducer leads to modification of the crystalline components of the AMR, shown in the figure by subtracting the curves at piezo voltages of $\pm 150$~V. This modification represents $\approx 10\%$ of the total AMR and is comparable to the absolute magnitude of the crystalline terms. The appearance of a fourth order term in the transverse AMR is expected under a uniaxial distortion \cite{to be published}, but this higher order term was not found to be significant in unstrained 25nm (Ga,Mn)As layers \cite{Rushforth:2007_a}.

\begin{figure}[t]
%\vspace{1cm}
\hspace{0cm}\includegraphics[width=0.7\columnwidth,angle=90]{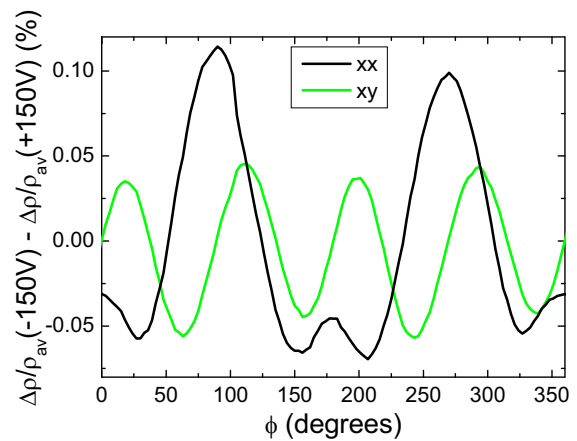}

%\vspace{0cm}
%\caption{\textbf{Figure1}}
\caption{The change in (a) the longitudinal $\Delta\rho_{xx}/\rho_{av}$ and (b) the transverse $\Delta\rho_{xy}/\rho_{av}$ components of the AMR for piezo voltages of $\pm 150$~V.}
\label{f4}
\end{figure}

To conclude, we have demonstrated the voltage control of the magnetic anisotropy and non-volatile switching of the magnetisation direction in (Ga,Mn)As induced by strain applied with a piezoelectric transducer. Microscopic theory calculations capture the physics involved. These techniques open up new avenues for exploring a variety of micromagnetic and magnetotransport phenomena in DMS systems.

%\newpage

\textbf{Acknowledgements}
We are grateful to J. Chauhan and D. Taylor for sample fabrication. We acknowledge support from EU Grant IST-015728, from UK Grant GR/S81407/01, from CR  Grants 202/05/0575, 202/04/1519, FON/06/E002, AV0Z1010052, KJB100100802 and LC510.

%\newpage

%\textbf{Author Information}
%The authors declare no competing financial interests.

%\newpage
%\textbf{References}
%\bibliography{MSWEBpublications}

\end{document}